\documentclass[doc]{apa6}

\usepackage[utf8]{inputenc}
\usepackage[american]{babel}
\usepackage{float}
\usepackage{csquotes}
\usepackage{array}
\usepackage{hyperref}
\usepackage{amsmath}
\usepackage{subcaption}
\usepackage{tabularx}
\usepackage{spverbatim}
\usepackage{tikz}
\usetikzlibrary{bayesnet}
\usepackage[style=apa,sortcites=true,sorting=nyt,backend=biber]{biblatex}
\DeclareLanguageMapping{american}{american-apa}
\title{The experiment is just as important as the likelihood in understanding the prior: A cautionary note on robust cognitive modelling.}
\shorttitle{Modelling and Cognitive Science}
\threeauthors{Lauren Kennedy*}{Daniel Simpson}{Andrew Gelman}
\threeaffiliations{Columbia Population Research Center and Department of Statistics, Columbia University}{Department of Statistical Sciences, University of Toronto}{Department of Statistics and Department of Political Science, Columbia University}

\abstract{Cognitive modelling shares many features with statistical modelling, making it seem trivial to borrow from the practices of robust Bayesian statistics to protect the practice of robust cognitive modelling. We take one aspect of statistical workflow---prior predictive checks---and explore how they might be applied to a cognitive modelling task. We find that it is not only the likelihood that is needed to interpret the priors, we also need to incorporate experiment information as well. This suggests that while cognitive modelling might borrow from statistical practices, especially workflow, care must be taken to make the necessary adaptions.}

\authornote{This is a comment on the article ``Robust Modeling in Cognitive Science'' \parencite{Lee2019} for Computational Brain and Behavior. We extend our thanks to Steven Langsford for his insightful comments on this manuscript, and Wai Keen Vong for earlier comments.  We also thank the U.S. Institute for Education Sciences
and Office of Naval Research for partial support of this work. \newline
*Corresponding author: lauren.kennedy729@gmail.com}

\begin{document}

\maketitle
Cognitive modelling aims to create generative, phenomenological models of cognitive processes. In contrast, statistical tools are mostly designed to disentangle a signal from measurement error in the data. This leads to some tension between the tasks statistical tools are designed for and the types of questions cognitive models attempt to answer. In this paper we highlight that it is important to remain aware of the difference in purpose when trying to apply modern (or, for that matter, classical) statistical techniques to cognitive modelling. We focus on Bayesian methods, but we believe that the general warning should hold more generally. 

We highlight this by exploring how an increasingly popular concept from statistical modelling---viewing modelling as a holistic workflow rather than a set of discrete and independent activities---can be applied to cognitive modelling practices.  This idea, as outlined in \textcite{gabry2019visualization,betancourt_workflow_cs,PrincipledCognitiveScience}, can be broken into two phases: pre-experiment prophylaxis and post-data model evaluation, criticism, and extension.   In the interest of brevity, this will focus primarily on the pre-data modelling phase.  In particular we wish to highlight the functionality of techniques like prior predictive checks for visualizing the impact of supposedly uninformative priors on the expected observed values. The consideration of priors is specifically of interest to Bayesian analysis but other simulation based techniques, such as simulation based power analyses are similarly useful for considering expected data. 

Prior predictive checks are a method to robustify Bayesian  cognitive modelling that was not covered in detail in the excellent paper of \textcite{Lee2019}. Furthermore, in preparing this paper we were struck that the differences between cognitive modelling and statistical modelling implies that we should be cautious when applying statistical tools for model evaluation to cognitive modelling. 

To achieve this we argue how the generative nature of cognitive models and our knowledge of the experiment being used to collect data should be used to inform and criticize our choices of prior.  We demonstrate that prior predictive checks can be adapted to provide a useful method for visualizing the impact of prior choices in cognitive models, which we demonstrate using the Balloon Analogue Risk Task \parencite[BART;][]{lejuez2002evaluation}. We argue that this technique provides information regarding prior specification over and above simulation studies of parameter recovery and model selection. 

By doing so we demonstrate an important challenge. Methods designed purely for use in statistical models (most often general linear models) cannot be directly adapted to the challenges faced by cognitive modelling without adaption. 

\section{Visualize priors and complex likelihoods with prior predictive checks}
Cognitive models are designed to postulate a generative process for the complex procedures that occur within the brain. Given this complexity, it is hardly surprising that this generative process is often far more complicated than most models proposed for purely data-driven statistical analyses. However, modelling of data and modelling of the brain do share some commonalities, and we believe these commonalities (e.g., the prevalence of a Bayesian framework, the desire to interpret particular parameters, and the need to compare models) suggest the potential application of robust statistical modelling practices within a cognitive modelling framework. 

\section{Example: Balloon Analogue Risk Task}

We frame the remainder of this comment in the context of an illustrative example, a task is designed to investigate risk aversion through the trade off between choosing to pump a balloon (increasing expected reward for a given trial provided pumping does not pop the balloon) and cashing out a trial. For simplicity, we use the model described in \textcite{lee2014bayesian} but originally proposed by \textcite{van2011cognitive} and reproduced in Figure~\ref{tab:models}, left panel, for reader convenience. 

We use the simplified model and experimental design described in \textcite{lee2014bayesian}, with their accompanying data for one paricipant who participates in $30$ trials of the balloon task in each of three levels of intoxication (sober, tipsy and drunk). The probability that balloon will burst on any given trial is given at the start of the trial and held constant throughout the trial. In the data we consider here the probability of popping is .10, .15 and .20 for 30 trials of each. The data we use (purely for illustrative purposes) is from \textcite{lee2014bayesian}. 

For now we focus on the non-hierarchical model in panel 1, but the hierarchical model explores the difference between sober, tipsy and drunk conditions. 

\begin{table}[h]
    \begin{subtable}[h]{0.17\textwidth}
        \begin{tabular}{c}
          \begin{tikzpicture}

  \node[obs]                               (p) {$p$};
  \node[det, below=of p, xshift=0cm]    (w) {$\omega   $};
  \node[latent, below=of p, xshift=-1.2cm] (g) {$\gamma^{+}$};
  \node[det,    below= of w]               (theta) {$\theta_{j,k}$};
  \node[latent, below= of w, xshift=-1.2cm] (b)  {$\beta$};
  \node[obs,    below= of theta]            (d) {$d_{j,k}$};

  \edge {p,g} {w} ;
  \edge {b,w} {theta} ;
  \edge {theta} {d} ;%

  \plate {plate1} {(theta)(d)} {$k$ choices} ;
  \plate {} {(plate1)(theta)(d)} {$j$ trials} ;

\end{tikzpicture}
        \end{tabular}
    \caption{Non-hierarchical model}
    \end{subtable}
    \begin{subtable}[h]{0.20\textwidth}
            \begin{tabular}{l}
              $\gamma^+ \sim $Uniform$(0,10)$\\
              $\beta \sim $Uniform$(0,10)$\\
              $\omega = -\gamma^+/$log$(1-p)$\\
              $\theta_{j,k} = 1/(1+$exp$(\beta(k-\omega))$\\
              $d_{j,k} \sim $Bernoulli$(\theta_{j,k})$
            \end{tabular}
       \label{tab:non_hier_m}
    \end{subtable}
    \hfill
    \begin{subtable}[h]{0.30\textwidth}
        \begin{tabular}{l}
            \begin{tikzpicture}

  \node[obs]                               (p) {$p$};
  \node[det, below=of p, xshift=0cm]    (w) {$\omega_i$};
  \node[latent, below=of p, xshift=-1.2cm] (g) {$\gamma^{+}_i$};
  \node[latent, left=of g]              (sg){$\sigma_{\gamma^+}$};
  \node[latent, above=of sg]            (ug){$\mu_{\gamma^+}$};
  \node[det,    below= of w]               (theta) {$\theta_{i,j,k}$};
  \node[latent, below= of w, xshift=-1.2cm] (b)  {$\beta_i$};
  \node[latent, left=of b]              (sb){$\sigma_\beta$};
  \node[latent, below=of sb]            (ub){$\mu_\beta$};
  \node[obs,    below= of theta]            (d) {$d_{i,j,k}$};

  \edge{ug,sg} {g};
  \edge{ub,sb} {b};
  \edge {p,g} {w} ;
  \edge {b,w} {theta} ;
  \edge {theta} {d} ;%

  \plate {plate1} {(theta)(d)} {$k$ choices} ;
  \plate {plate2} {(plate1)(theta)(d)} {$j$ trials} ;
  \plate {} {(plate2)(g)(b)(w)(theta)(d)} {$i$ conditions} ;

\end{tikzpicture}
        \end{tabular}
        \caption{Hierarchical model}
        \label{tab:hier_m}
    \end{subtable}
    \begin{subtable}[h]{0.20\textwidth}
        \begin{tabular}{l}
            $\mu_{\gamma^+} \sim $Uniform$(0,10)$\\
            $\sigma_{\gamma^+} \sim $Uniform$(0,10)$\\
            $\mu_\beta \sim $Uniform$(0,10)$\\
            $\sigma_\beta \sim $Uniform$(0,10)$\\
            $\gamma^+ \sim N(\mu_{\gamma^+},\sigma_{\gamma^+})$\\
            $\beta \sim N(\mu_\beta,\mu_\sigma)$\\
            $\omega = -\gamma^+/$log$(1-p)$\\
            $\theta_{j,k} = 1/(1+$exp$(\beta(k-\omega))$\\
            $d_{j,k} \sim $Bernoulli$(\theta_{j,k})$
        \end{tabular}
    \end{subtable}
     \captionof{figure}{Non-hierarchical and hierarchical models of the Balloon Analogue Risk Task. Reproduced from \textcite{lee2014bayesian}, including using uniform priors, which would not be our first choice.}
     \label{tab:models}
\end{table}

In this model the observed parameters are $p$, the probability the balloon will pop, and $d$, the decision made by the individual on a given trial (i.e., pump or take reward). The remaining parameters control the expected number of pumps ($\beta$) and the between trial variation ($\gamma^+$). 

\subsection{Prior Predictive Checks}

Previous advice on specifying priors has encouraged the use of ``uninformative priors'' to prevent researcher bias (through prior selection) from impacting the model. However, work by \textcite{gabry2019visualization} demonstrates that uninformative priors, when interpreted in terms of the likelihood can actually be very informative. In fact, \textcite{gelman2017prior} argue that ``The prior can often only be understood in the context of the likelihood'', which is already filtering through to cognitive science \parencite{PrincipledCognitiveScience}, albeit in contexts where mixed effects (or multilevel) models are appropriate. While we use this technique in an exploratory manner in this manuscript, there is no reason why it couldn't be employed in a confirmatory manner. 

However, in cognitive modelling, we argue that more is needed than the likelihood to understand the prior. We draw the reader's attention in particular to the uniform hyper priors at the highest level of both non-hierarchical and hierarchical versions of the model. The use of uniform priors at the highest level of the hierarchy is common as it is assumed that these parameters will be well informed by the observed data and hence need little prior regularization. However,  diffuse priors on parameters on lower levels of the hierarchy can be  surprisingly informative once they filter through the likelihood \parencite{gabry2019visualization, gelman2017prior,PrincipledCognitiveScience}. One technique to understand the priors in terms of the likelihood are prior predictive checks. 

Prior predictive checks work by recalling that a Bayesian model is a generative model for the observed data (cf. a cognitive model, which is a generative model for an underlying process). This implies that simulations from the full Bayesian model (that is simulating parameters from the prior, simulating cognitive behaviour from the cognitive model, and then simulating measurements from the observation model) should be plausible data sets.  For example, they should not imply that there is a low probability that a drunk undergraduate on their 20th trial will pump the balloon less than 50 times.  In this way, prior predictive checks can be thought of as sanity checks that the prior specification and observation model have not interacted badly with the cognitive model.

Two things can go wrong when the prior predictive distribution does not conform to our substantive understanding of the experiment. The lesser problem is that these models can cause problems with the computational engine being used for inference. This is a manifestation of the ``folk theorem'' of statistical computing: when  you have computational problems, often there’s a problem with your model \parencite{FolkThm}. 

The more troubling problem that can occur when the prior predictive distribution puts a lot of weight on infeasible parts of the data space is that this will affect inference. In extreme cases, any data that is observed will be in conflict with the model, which will lead to poor finite-sample performance. Even in the situation where we know that asymptotically the inference will be consistent, the data will still use up a lot of its information in overcoming the poor prior specification. This will lead to underpowered inferences.
 
 Prior predictive checks should complement the already prevalent practice of parameter recovery checks. These  empirically assess the identifiability of model parameters by simulating data from the model using known parameters and checking that the posterior recovers these parameters when this particular experimental design is used.  \textcite{talts2018validating} show that a combination of prior simulations and parameter recovery checks can be used to validate inference software.

\begin{figure}[h]
 \begin{subfigure}{0.5\textwidth}
\includegraphics[width=0.9\linewidth]{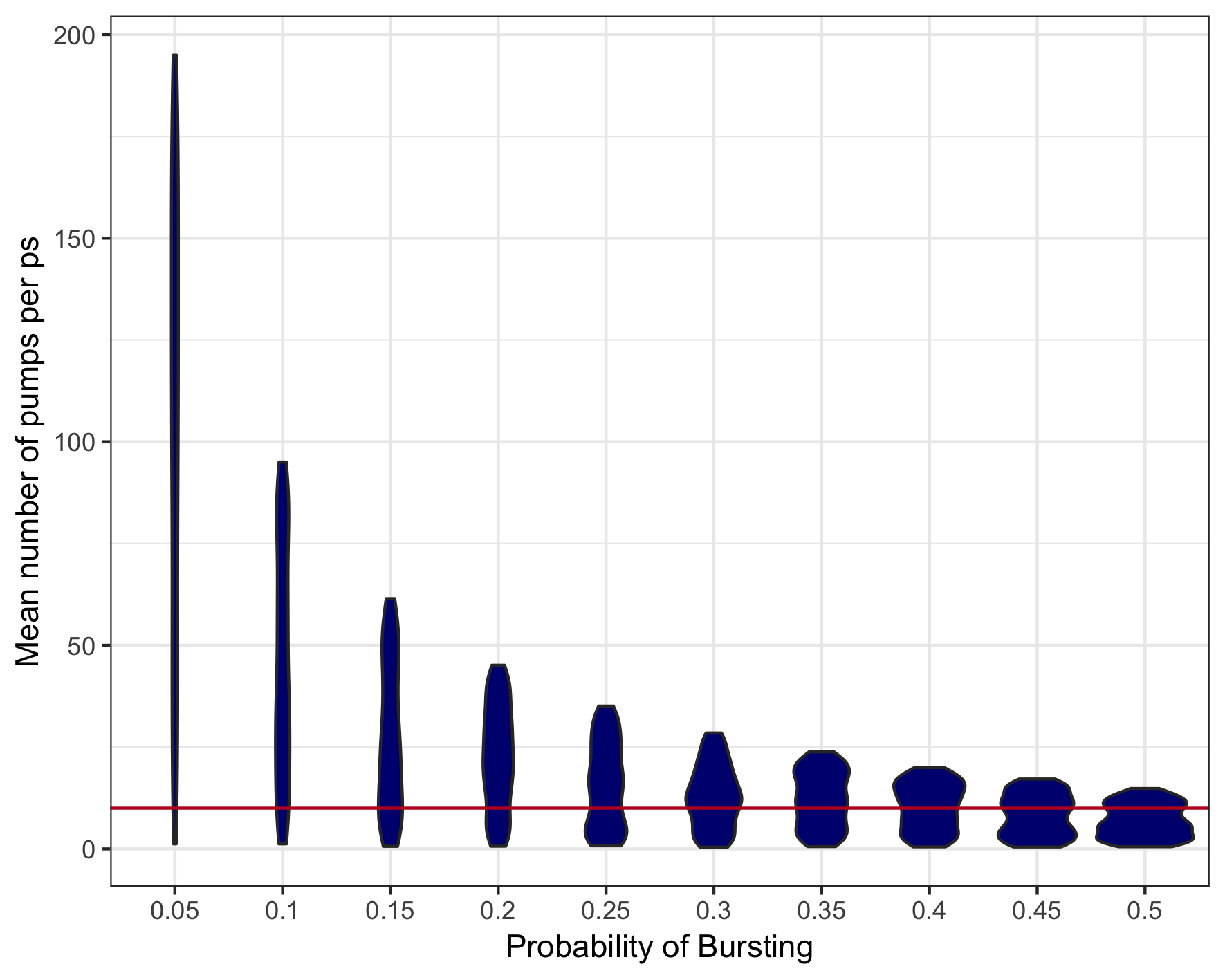}
\caption{Mean number of pumps using prior checks with likelihood, prior.}
\label{fig:simple_bart_ppcs_sd}
\end{subfigure}
\begin{subfigure}{0.5\textwidth}
\includegraphics[width=0.9\linewidth]{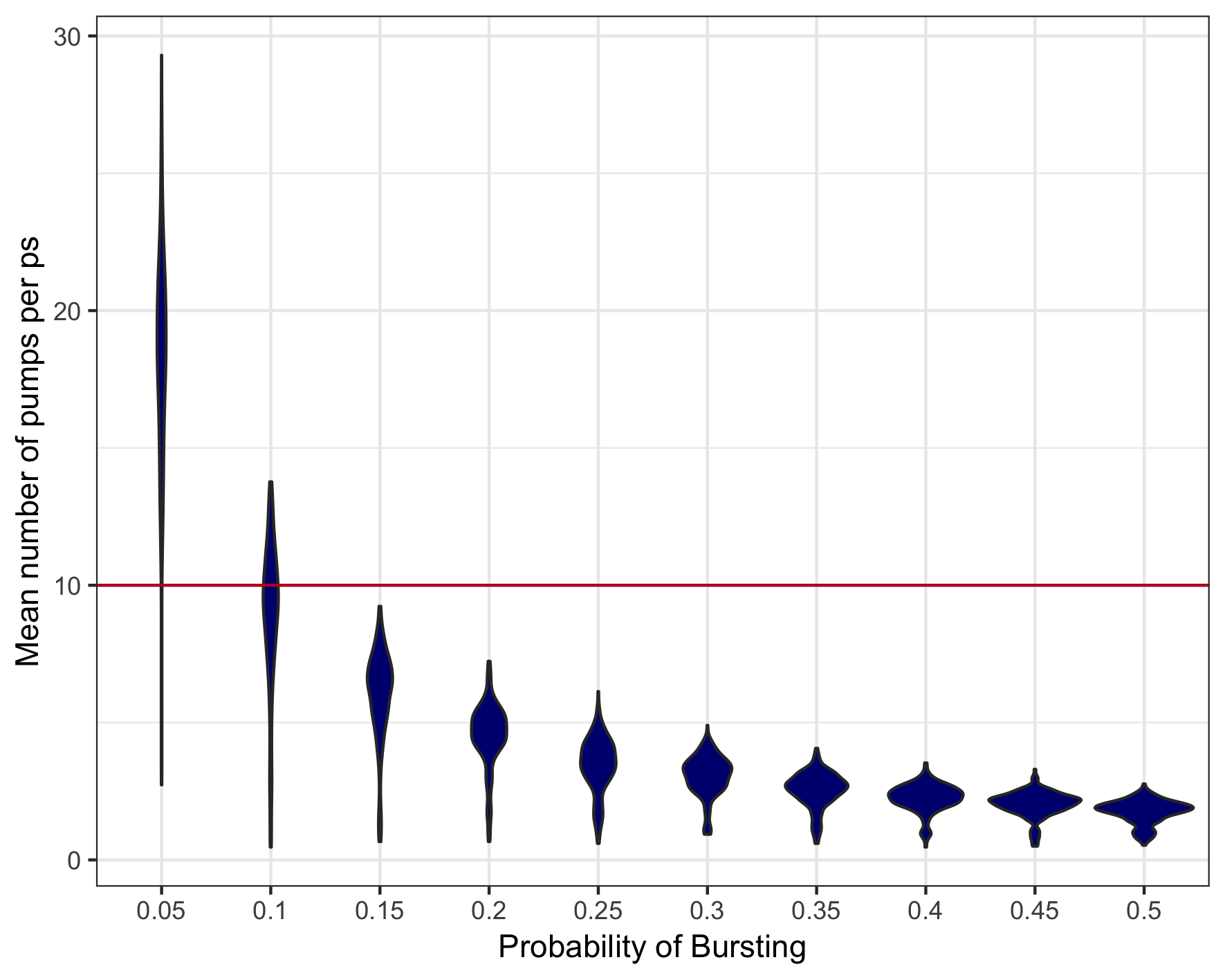}
\caption{Mean number of pumps using prior checks with likelihood, prior  and experimental design.}
\label{fig:simple_bart_ppcs_mean}
\end{subfigure}

\caption{Prior checks with likelihood and likelihood plus experimental design. Here we simulated the expected number of pumps for 200 participants with different probabilities of the balloon popping (x-axis), and plot the average number of pumps per participant on the y axis. Note that while both prior checks suggest increasing the probability of the balloon popping decreases the expected number of pumps, the majority of simulated pumps are exceptionally high for the prior checks using only the prior and likelihood (left panel), but not when using the experimental design as well (right panel), and the shape of the expected distribution is markedly different.}
\label{fig:simple_bart_ppcs}
\end{figure}

\subsubsection{Non-hierarchical model}

To demonstrate this, we simulate values from the priors for the non-hierarchical model for the balloon task and then use these priors to predict the expected outcome---in this case the number of pumps---using the likelihood. We simulate each trial as an independent participant to show the distribution of expected number of balloon pumps. In Figure~\ref{fig:simple_bart_ppcs} we compare this number to the number of pumps by participant George (data available from \textcite{lee2014bayesian}), which reflects the number of pumps observed in published literature from this task. The prior checks suggest that the number of pumps could reach up to $200$ expected pumps, but the observed number of pumps do not extend beyond $10$ (marked with red line). 

However, in cognitive modelling there is more to interpreting the priors than merely the likelihood. In our example part of the experimental design is that after each decision, there is some probability that the balloon pops, ending the trial. This is not included in the likelihood, but is a feature of the experimental manipulation. When we incorporate this into the prior predictive checks---adding to the likelihood---we see that the expected number of pumps is much more reasonable. 

How can we say that the prior expected number of pumps is reasonable? The natural response to this is to say that before the experiment was conducted, we could not possibly have known the expected number of times the participants would choose to pump the balloon. Although often cognitive modelling examples are not as clear as that proposed by \textcite{gabry2019visualization}, where the priors suggested a pollution concentrate so thick that human life could not survive, experimental experience suggests that a participant is exceptionally unlikely to commit to pressing a button 200 x 90 times for limited increase in payoff (or a potential decrease in payoff) in various stages of inebriation. More recommendations on prior selection for cognitive modelling can be found in \textcite{lee2018bayesian,lee2018determining}.

Using these adapted prior checks, we can determine that while the wide uniform priors might look like they could potentially be informative, when combined with the experimental design they do not suggest that the number of pumps will be exceptionally high. In Figure \ref{fig:hier_BART_ppcs}, we see that increasing the width of the uniform prior only changes the length of the tail of the distribution in the number of pumps. 

Not only do prior predictive checks help to understand the priors, they also help us to understand the informativeness of data to distinguish between different values of parameters. For example, the prior width on both parameters extend above $5$ suggests only slight differences in the tails of the distribution of the expected number of pumps. If it is truly expected that the number of pumps have an impact in the tails, then considerable data will be needed to observe this difference. 

\subsubsection{Hierarchical model}

\begin{figure}[h]
\begin{subfigure}{0.5\textwidth}
    \includegraphics[width=0.9\linewidth, height=5cm]{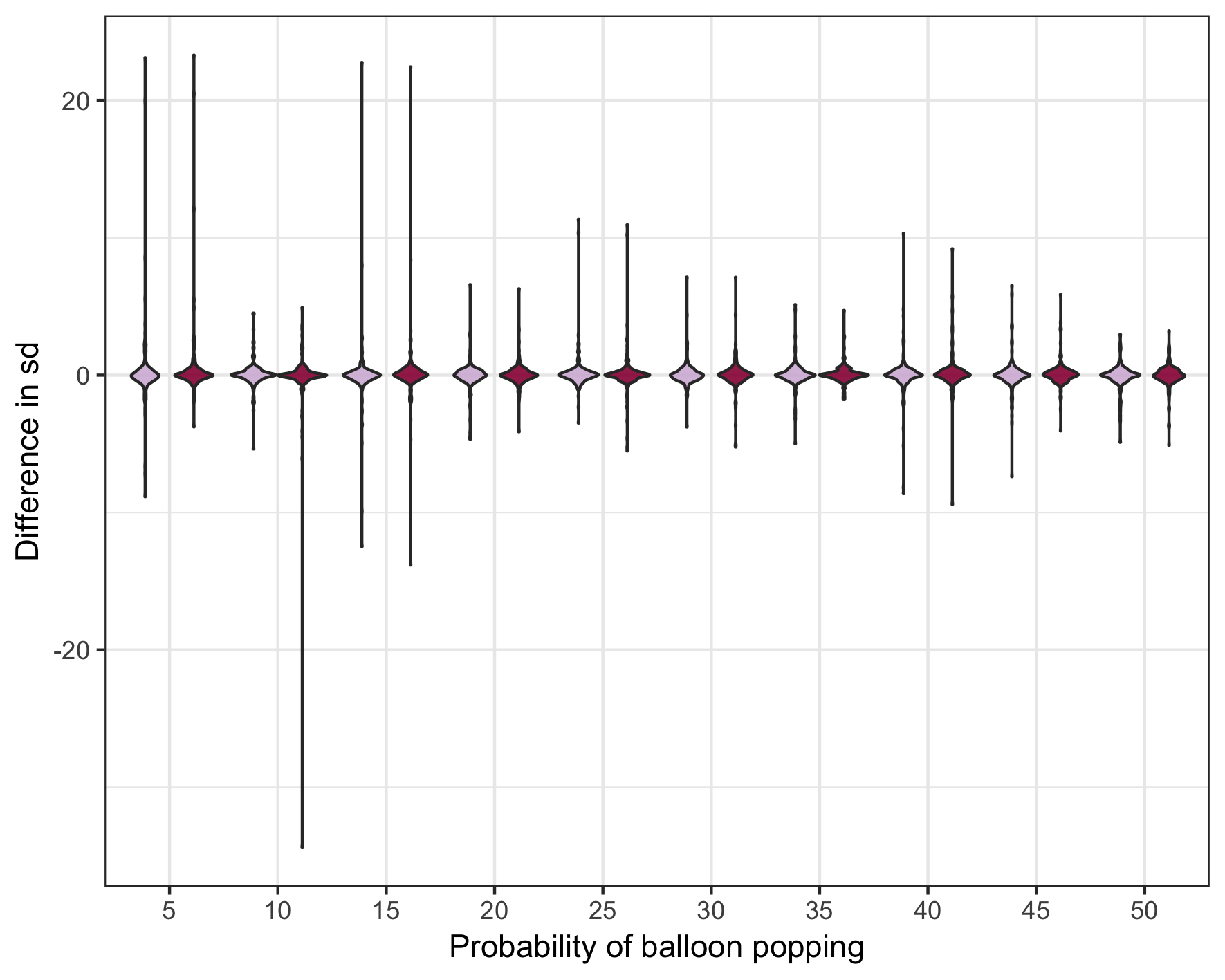}
    \caption{Difference in mean from simulated sober condition}
    \end{subfigure}
    \begin{subfigure}{0.5\textwidth}
    \includegraphics[width=0.9\linewidth, height=5cm]{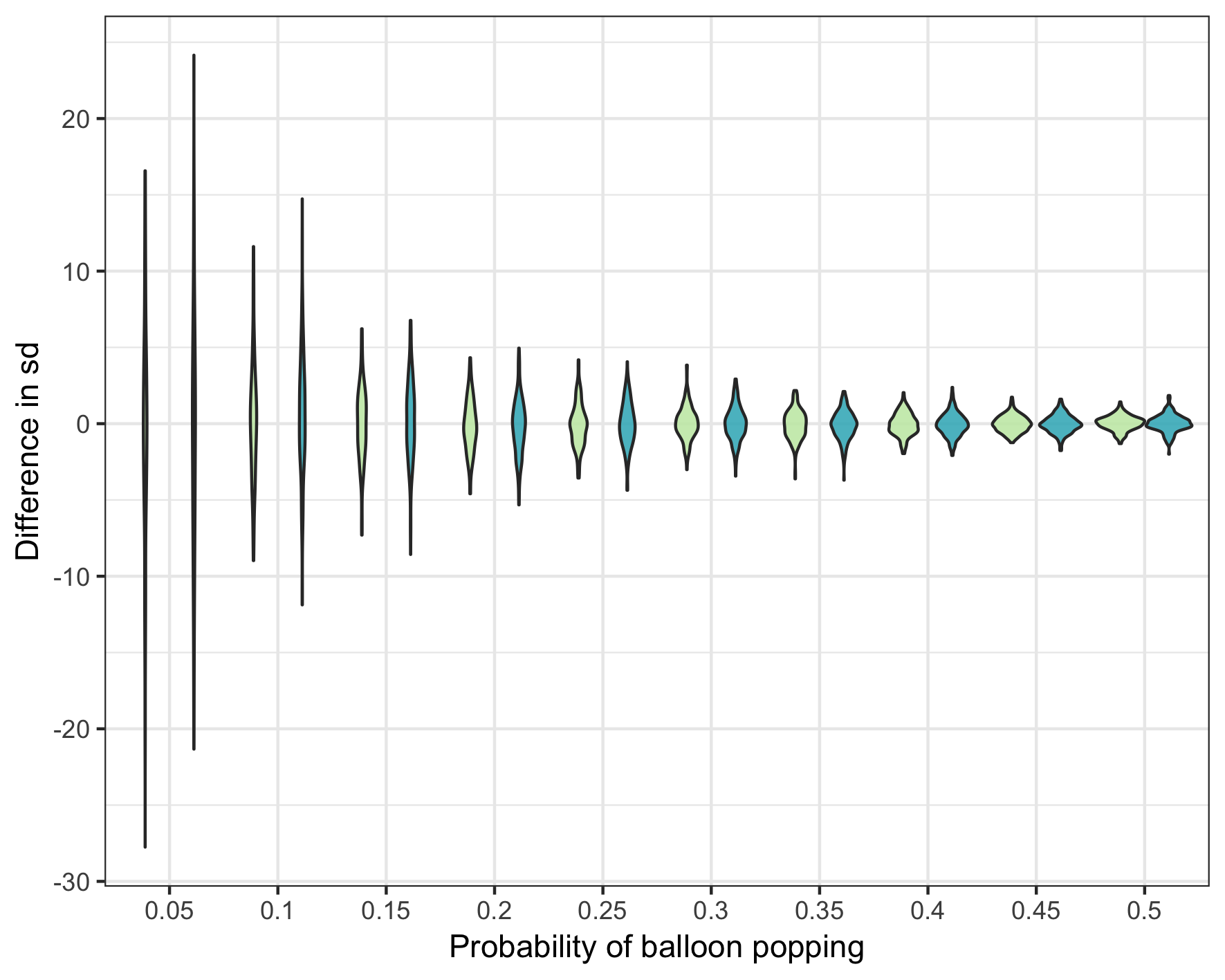}
    \caption{Difference in mean from simulated sober condition}
    \end{subfigure}
\begin{subfigure}{0.5\textwidth}
    \includegraphics[width=0.9\linewidth, height=5cm]{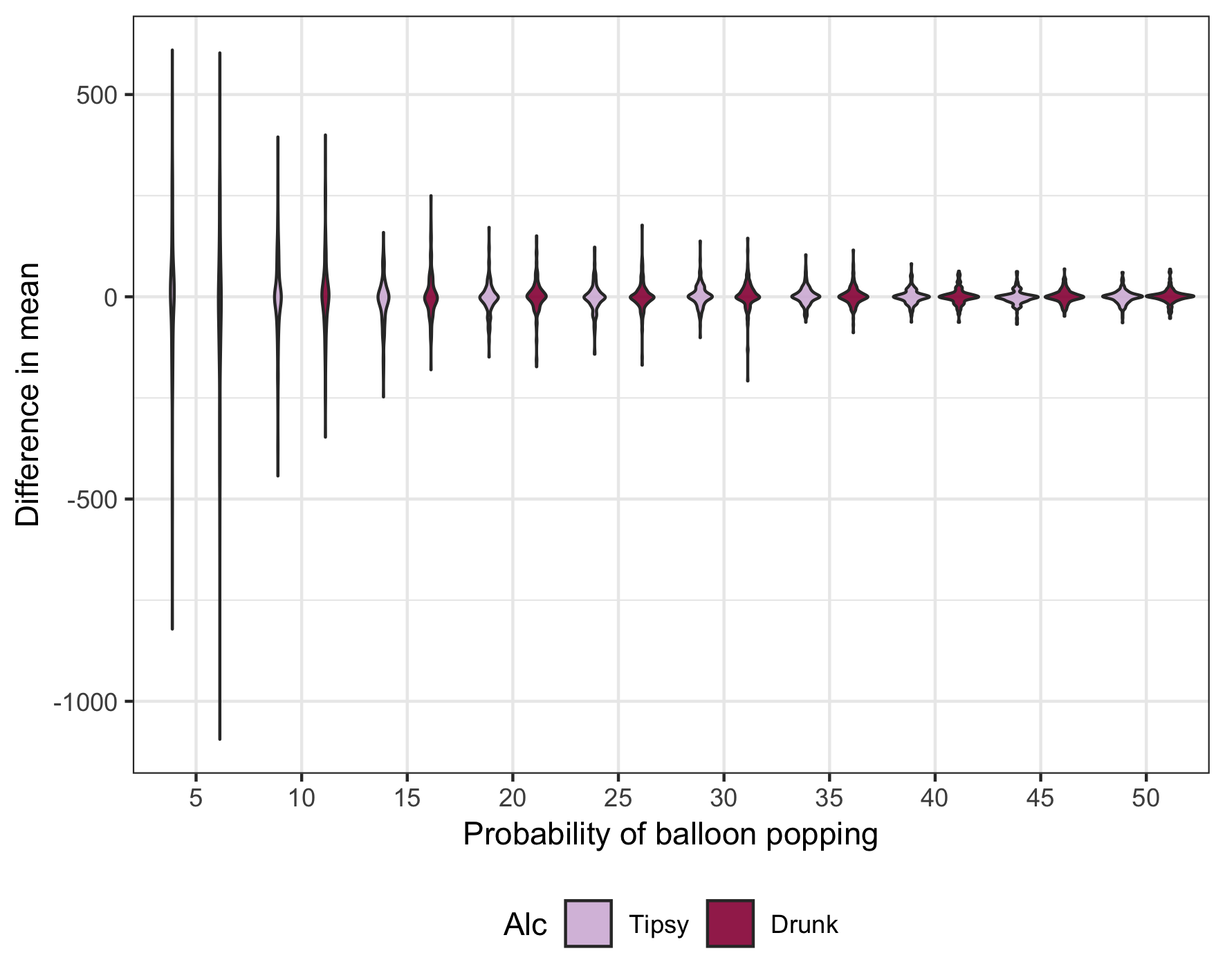}
    \caption{Difference in variance from simulated sober condition}
\end{subfigure}
\begin{subfigure}{0.5\textwidth}
    \includegraphics[width=0.9\linewidth, height=5cm]{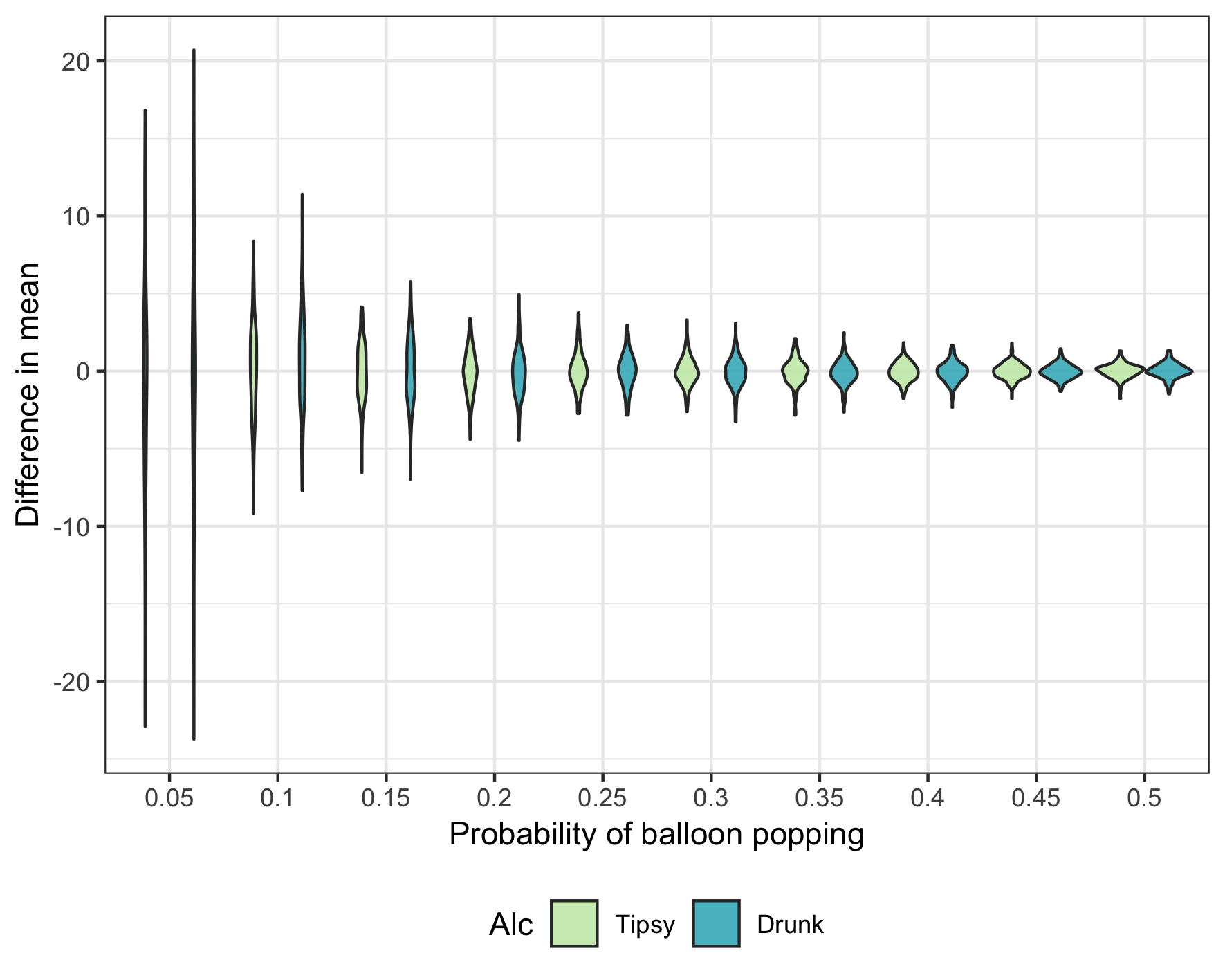}
    \caption{Difference in variance from simulated sober condition}
\end{subfigure}
\caption{Comparison of how the width of the uniform impacts the expected difference between conditions in both variance (top) and mean (bottom) using traditional prior checks (left), and prior checks using experimental design (right).}
\label{fig:hier_BART_ppcs}
\end{figure}

We see again that prior predictive checks that also incorporate the experimental design are preferred when we consider the hierarchical model. In Figure~\ref{fig:hier_BART_ppcs} we plot the difference in mean and variance between a simulated sober condition and the two simulated intoxicated conditions. Without modifying the prior checks, the hierarchical version of this model suggest that the expected variance of the number of pumps across trials is remarkably small. This suggests that once a participant choose a number of times to pump a balloon, they tend to commit to this number strongly across trials. Combined with the large expected number of trials this would be quite remarkable behaviour! However, when we consider the prior checks that incorporate experimental information, the design seem much more reasonable. 

\subsubsection{The difference between prior predictive checks for data modelling and cognitive modelling}

The first time we applied the prior predictive check idea to the model we did not take into account the fact that an experiment would sometimes end with the balloon popping. This meant that we drew incorrect conclusions about  the suitability of the priors. In hindsight, this is a consequence of the likelihood principle, which says that only the parts of the generative model that depend on the unknown parameters are needed in the likelihood in order to perform inference. However, the full generative model is needed to make predictions and, we argue, is also needed to do pre-data modelling. In this case, the probability of popping at each stage is a fixed, known parameter that is independent of all of the other parameters and is hence not in the specification given by \textcite{lee2014bayesian}, which means that using prior predictive checks directly from their model specification would incorrectly lead us to conclude that the priors are quite unreasonable.  

This turns out to be the fundamental challenge when adapting prior predictive checks to cognitive modelling. While in data analysis the outcome we want to predict is usually obvious from the structure of the model, in cognitive modelling it often requires further information about the context of the experiment.  In order to critique the model we need to know that the balloon will sometimes pop. In order to fit the model it is not strictly necessary to know this.

Unlike the efforts of \textcite{veksler2015model}, we do not mean to imply that prior predictive checks can be used for model selection. Rather, we argue that they highlight what data we could actually expect given the prior. In particular, they highlight the diminishing returns on the upper limit given the experimental design. They also give a sense of the power of the experimental design – if the experimenter is truly interested in differences larger than in parameter larger than 1, prior predictive checks suggest that an enormous amount of data would be needed as the observations would be very rare. Lastly they help the experimenter to select and quantify priors with complex likelihoods. 

\section{Model comparison}
One concern with potentially informative priors is that there is a carry through impact on the reliability of model comparison techniques. We believe that in some models this could be an unintended consequence, but for our balloon example we find that there are few differences when we simultaneously vary the width of the uniform priors included in the model. This further suggests that modifying the prior predictive checks technique has been useful. We use the George data included in \textcite{lee2014bayesian} and the Stan code included with this chapter. We made a few adjustments to increase computational efficiency (non-centered parameterization, use of combined functions), modified the code slightly to model different probability conditions, and made a few changes to ensure compatibility with the bridge sampling R package. Our final code is included on LK's Github page\footnote{https://github.com/lauken13/Comment-Robust-Cognitive-Modelling-}. We also randomly permuted the George data between intoxication conditions to investigate the evidence for the null. 

\subsection{Bayes factors}

\begin{figure}
\begin{subfigure}{0.5\textwidth}
\includegraphics[width=0.9\linewidth]{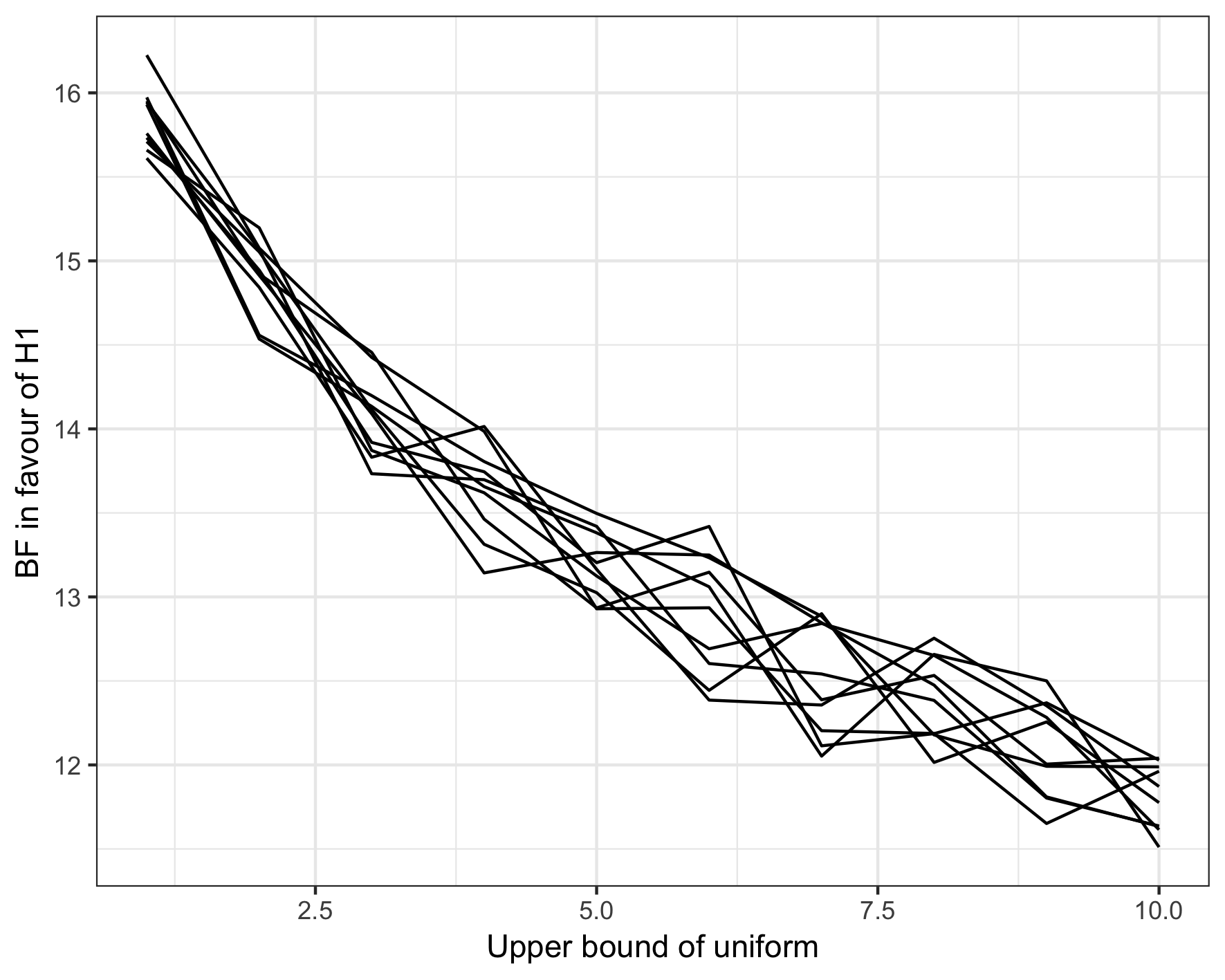}
\caption{George data}
\label{fig:hier_BF}
\end{subfigure}
\begin{subfigure}{0.5\textwidth}
\includegraphics[width=0.9\linewidth]{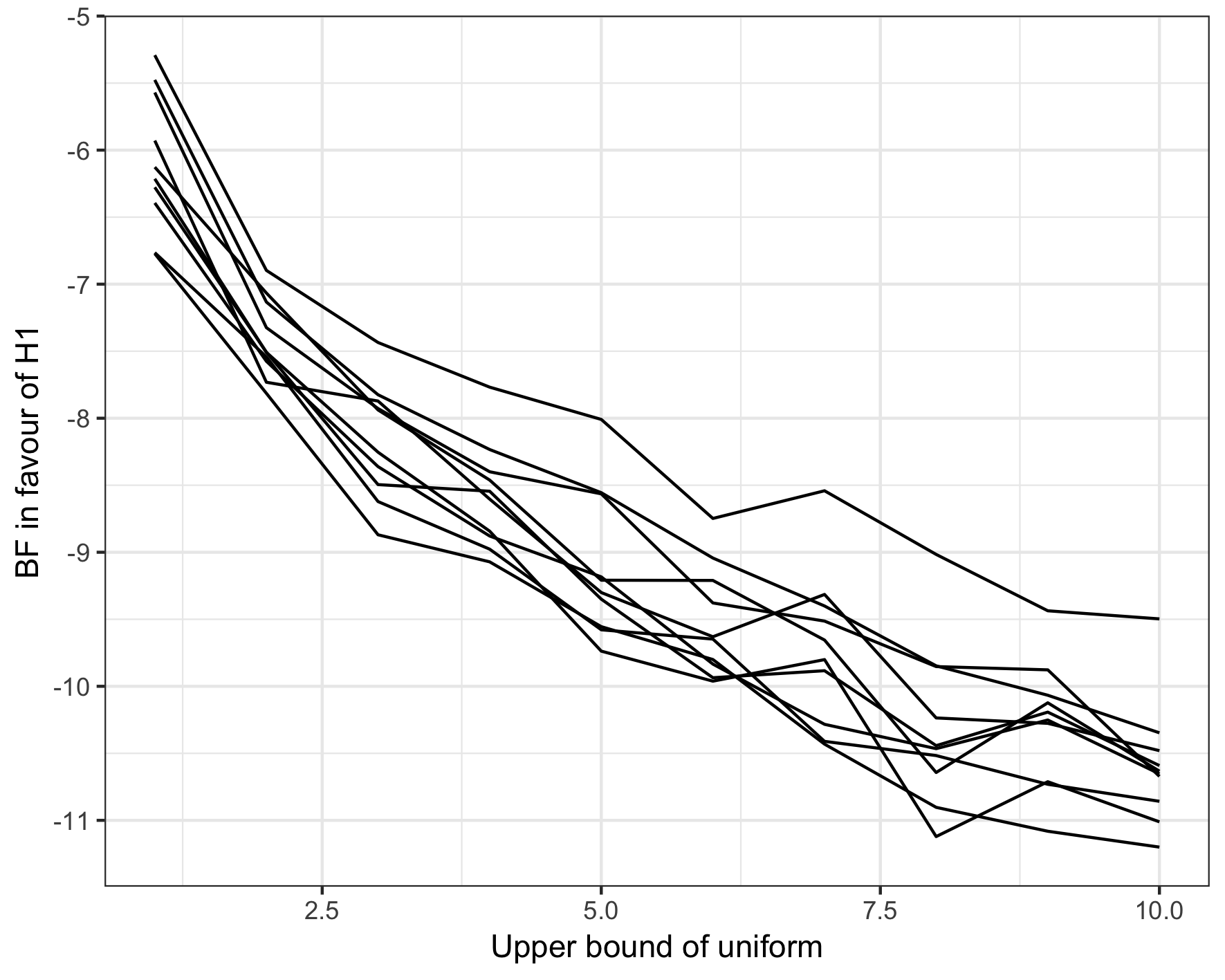}
\caption{Randomly permuted George data.}
\end{subfigure}
\caption{Comparison of the width of the uniform prior on the log Bayes factor for a hierarchical (H1, positive Bayes factor) against a non-hierarchical model (H0, negative Bayes factor).}
\label{fig:BF}
\end{figure}

We use the bridgesampling package \parencite{bridgesampling} to calculate Bayes factors for the hierarchical model when compared to the non-hierarchical model. As we can see in Figure~\ref{fig:BF}, while the evidence for the alternative reduces with priors of increasing width, the Bayes factor reliably suggests support for the alternative (left panel). We find similar results with permuted datasets (right panel). The Bayes factor does decrease with an increase in the widths of the uniform priors, but never so much as to suggest the hierarchical is preferred when the non-hierachical is true (right panel) or to change the conclusions from the George data (left panel).

\begin{figure}
\begin{subfigure}{0.5\textwidth}
\includegraphics[width=0.9\linewidth]{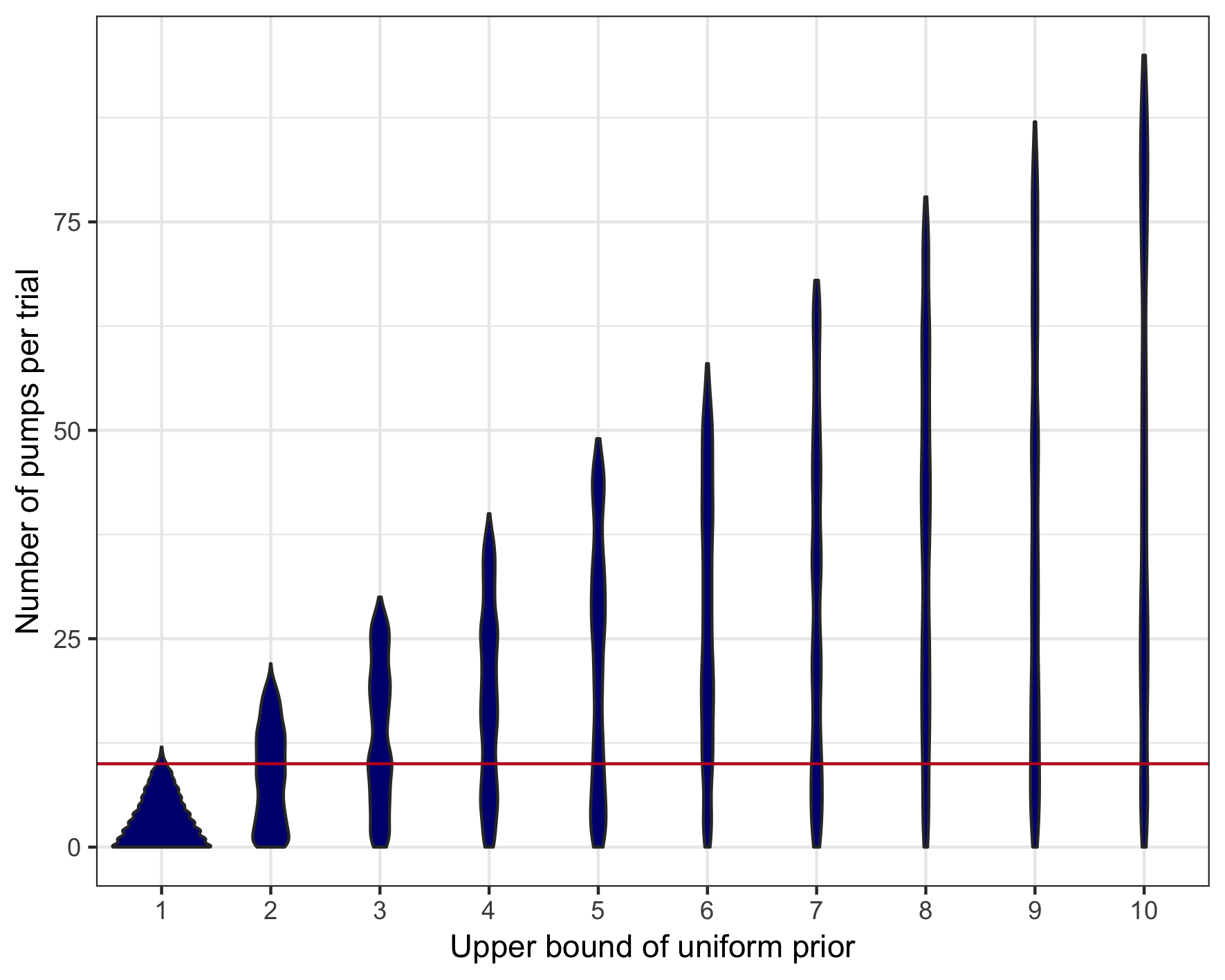}
\caption{Prior predictive check}
\end{subfigure}
\begin{subfigure}{0.5\textwidth}
\includegraphics[width=0.9\linewidth]{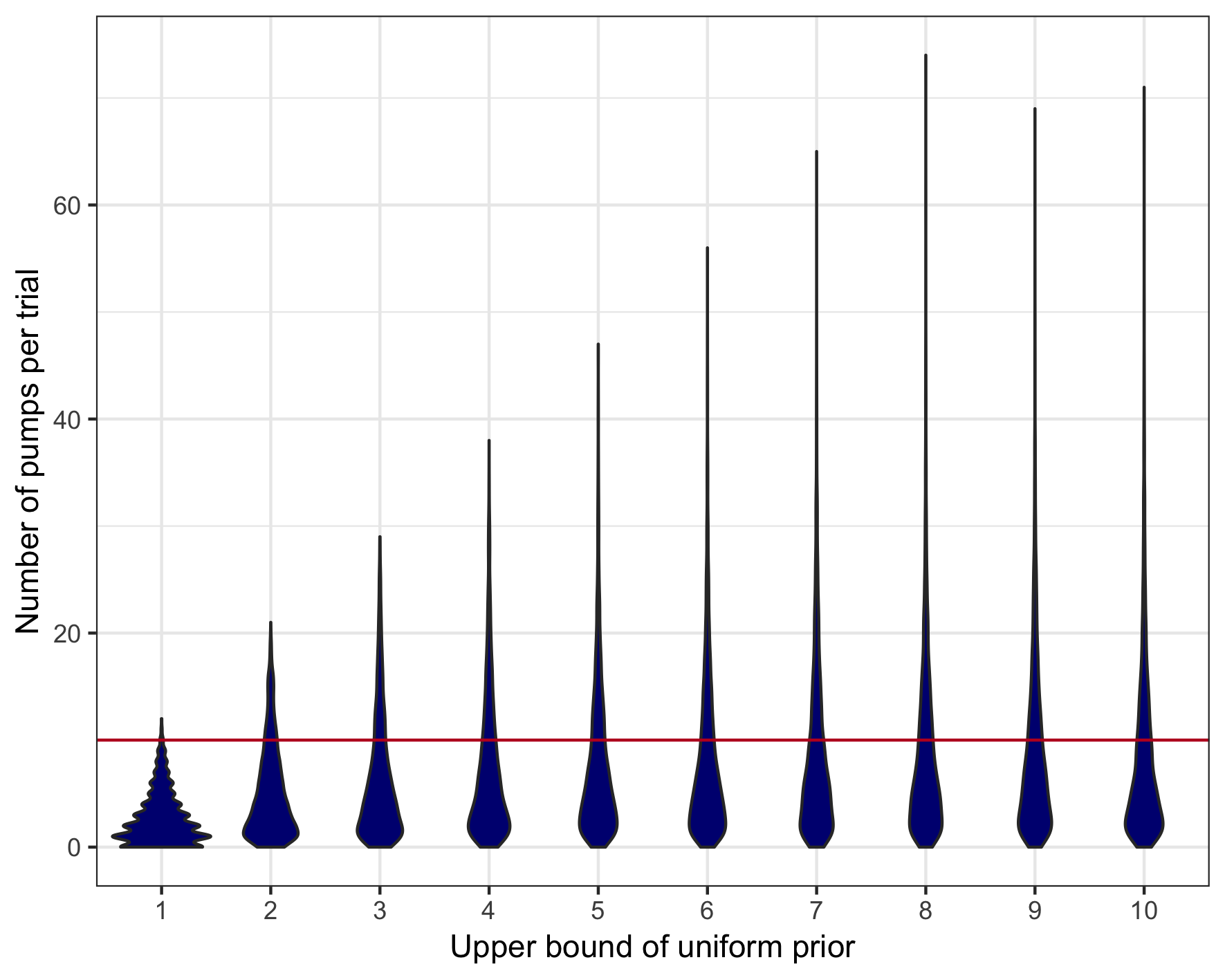}
\caption{Prior predictive checks with experimental design}
\end{subfigure}
\caption{Comparison of the width of the uniform prior (x axis) on the expected number of pumps (y axis) when the probability of the balloon popping is held constant at .10.}
\label{fig:upper_prior}
\end{figure}

The practical stability of the Bayes factors for this problem is related to the relative insensitivity of the prior predictive distribution to the upper bound on the uniform prior. Figure~\ref{fig:upper_prior} shows that  increasing this upper bound only slightly changes the tail of the predictive distribution. This is a demonstration of the principle that it is not so much that prior on the parameter that controls the behaviour of the Bayes factor, as the way that prior distribution pushes through the entire model.  In this case, a quite large change in the prior on a deep parameter like $\mu_{\gamma^+}$ only results in a mild change in the tail of the prior predictive distribution.

\subsection{Leave-one-out cross validation}

\begin{figure}
\begin{subfigure}{0.5\textwidth}
\includegraphics[width=0.9\linewidth]{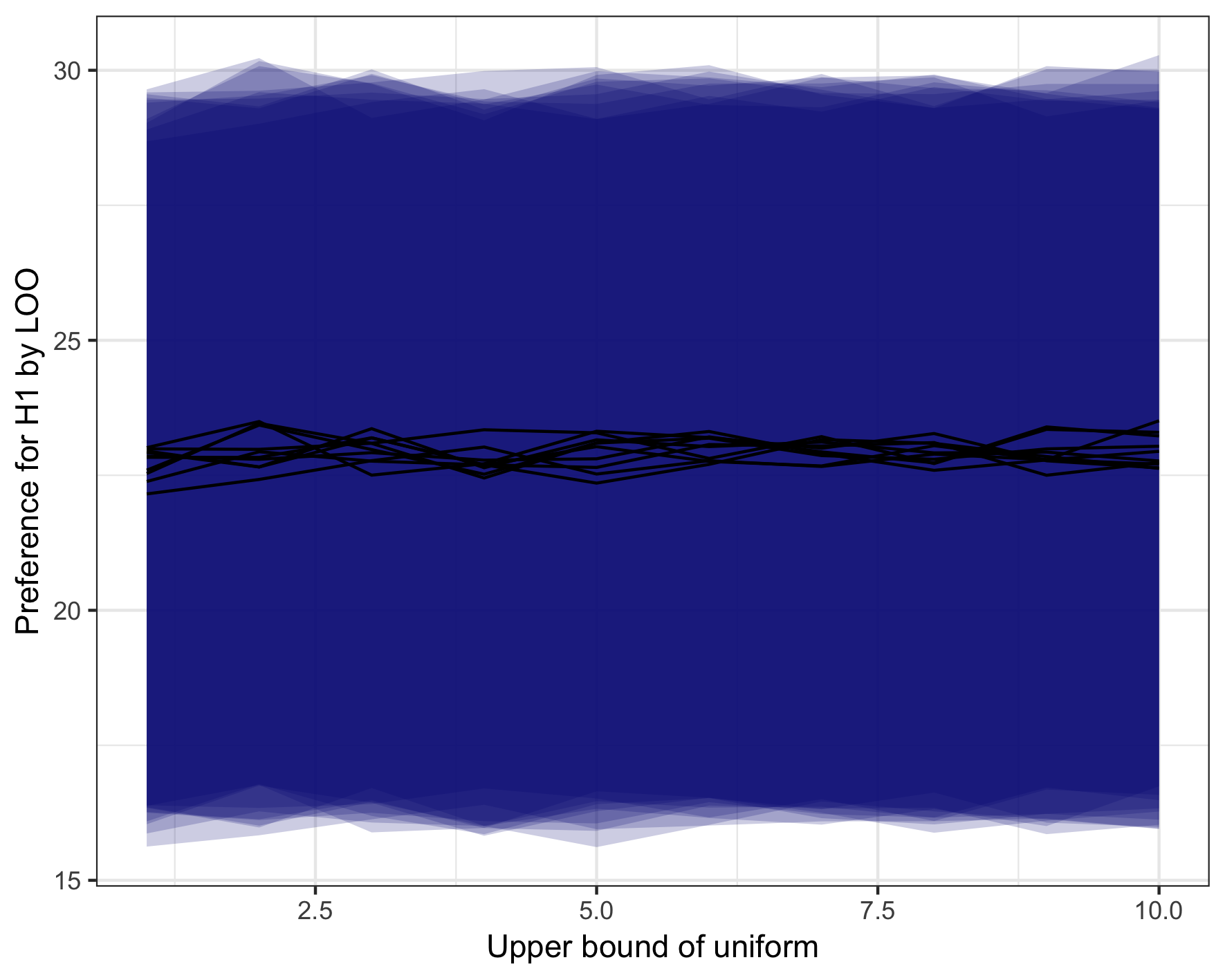}
\caption{George data}
\end{subfigure}
\begin{subfigure}{0.5\textwidth}
\includegraphics[width=0.9\linewidth]{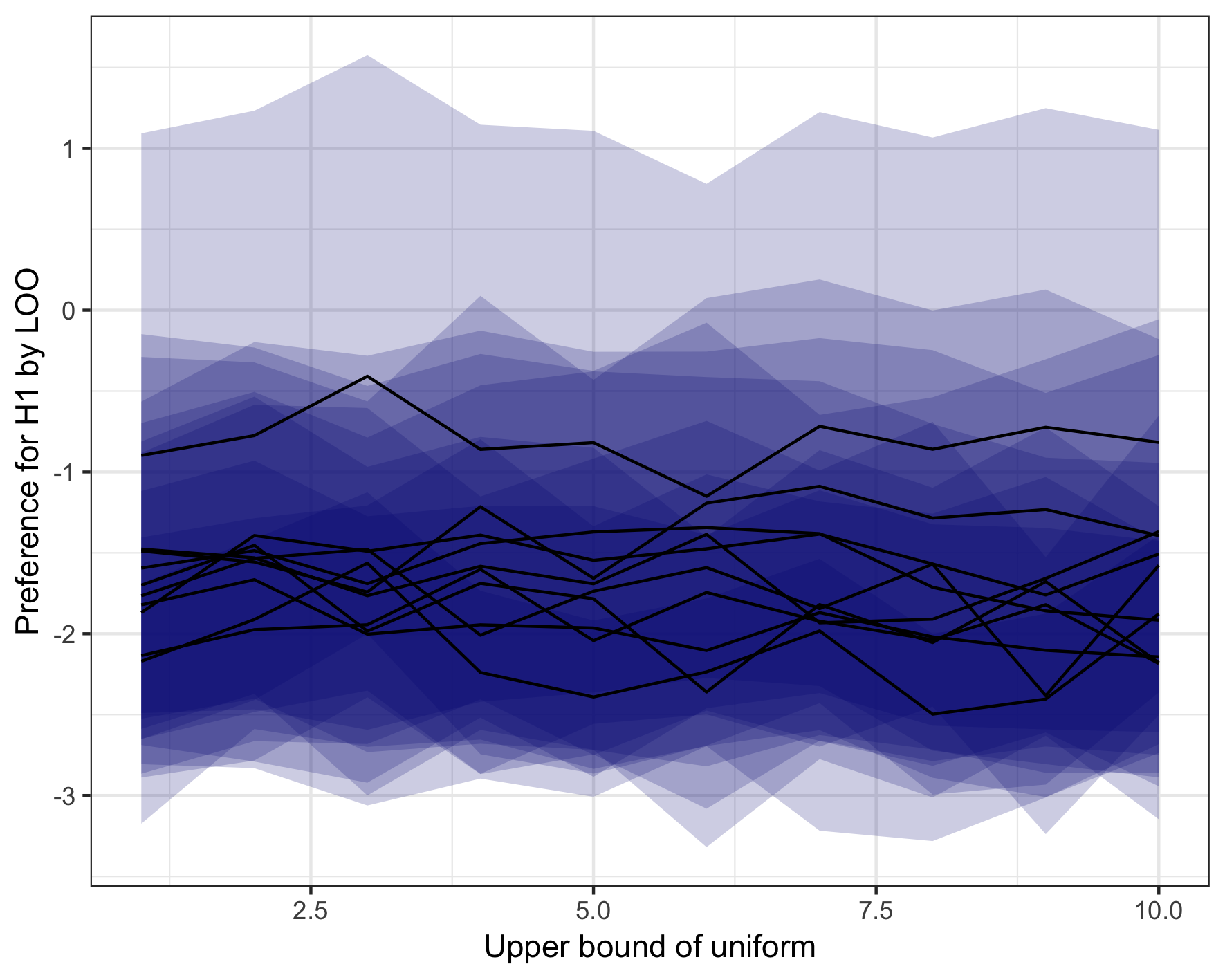}
\caption{Randomly permuted George data.}
\end{subfigure}
\caption{Comparison of the width of the uniform prior using LOO for a hierarchical (H1) against a non-hierarchical model (H0). Ribbons indicate the standard error of the ELPD approximation. }
\label{fig:LOO}
\end{figure}

As an alternative to the Bayes factor approach, we also employ an approximation to leave-one-out (LOO) cross validation using the LOO package \parencite{LOO}. For simplicity we leave one observation out (i.e., one choice to either pump or cash out), but more appropriate uses of LOO would leave one trial or condition out. As we can see in Figure~\ref{fig:LOO}, LOO estimates remain consistent over priors of increasing width.

This is in line with leave-one-out cross validation being sensitive to changes in the posterior predictive distribution. This is essentially unchanged by the width of the uniform priors.

\subsection{What constitutes a meaningful difference?}

Herein lies the problem with model comparison---if we are comparing a difference between conditions, are we hypothesizing that there is a difference between conditions for either $\gamma^+$ or $\beta$. If we were to test each separately we would need to compare a null model against a model with $\gamma^+$ varying, a null model with beta varying, and a model with both varying. With small $n$ and noisy data, what does it actually mean if we can distinguish between these models?

Given the relative flexibility of the modelling and the small range of potential scores in the actual observed data, is it possible to distinguish between small changes in $\gamma^+$ while holding $\beta$ constant and vice-versa? Another way of saying this to question whether the small effects in parameter estimates are due to sampler noise (MCMC error), measurement error, or due to actual differences in processes. 

To the extent that these questions can ever be resolved, we believe that prior simulation from the generative model \emph{for the data} has a role to play. They can be used to answer questions about what type of difference between models can be detected with the data at hand (we are deliberately avoiding the word \emph{power} here because this remains a vital question even outside the Neyman-Pearson framework).  Furthermore, simulation studies can and should be used to assess how these tools perform under different types of data model mis-specification. cognitive modelling can not be robust unless measurement error and mis-specification are properly considered.

\section{Statistical tools can't tell us what we want in practice. }
We've shown how prior predictive checks need to be adapted to understand the practical implications of priors in the context of cognitive modelling. Similarly model comparison tools don't tell us what we are most interested in. Regardless of method, they are all about prediction---predicted performance on the next participant provided they are exchangable with ones we have seen previously, predicted performance on the next iteration given nothing has changed from the previous iterations, prediction from the sample to the population assuming the sample is representative. Model comparisons tools are suited (although not all at once) to ask these questions,  but they all assume some type of equivalence. 

As \textcite{navarro2019between} notes, cognitive modelling asks a bolder question. Rather than prediction, we are often interested in extrapolation. Extrapolation to different participants, extrapolation to changes in condition, extrapolation to a population that is markedly different from our young and educated sample. We are interested not in whether subject George is less likely to pump a balloon given inebriation, but rather whether inebriation causes some cognitive change in \textit{people} that is realized in a \textit{general} risk aversion that is \textit{expressed} in a number of different domains. Moreover, as \textcite{navarro2019between} further notes we cannot answer these claims with a single experiment, not should we expect to answer these claims with any statistical analysis on a single experiment. 

Although this idea is controversial \parencite[see][for counter arguments]{evans_2019,Gronau2019}, we note that the argument made by \textcite{navarro2019between} calls for greater investment in generalization and extrapolation. Some ideas behind this are discussed in \textcite{Kennedy_MRP_psych}, but build upon much older ideas \parencite{shavelson1989generalizability}. 

We started this comment by claiming that cognitive models are similar to statistical models with greater interpretability and complexity. If this were true then robust cognitive modelling should borrow heavily from statistical sciences. However the work we present in this comment suggests that we cannot blindly apply the practices of statistics to cognitive modelling. Cognitive models share many traits of statistical models, and so we should employ prior predictive checks, model comparison tools and consider in sample prediction, but we need to do so with adaptions. 

We shouldn't expect to fall back on traditional statistical modelling tools but instead we should strive to reach further. \textcite{Lee2019} point to some avenues, but the reality is many tools are lacking because the basic assumptions like independence of the likelihood and the notion of equivalence held so dear to statisticians are the very assumptions that cognitive modellers want and need to violate. We find these challenges suggest a grand and exciting future for robust cognitive modelling and look forward to what the future will bring. 

\printbibliography{}

\end{document}